\documentclass[a4paper,12pt]{article}
\usepackage{cite}
\usepackage{color}

\usepackage{geometry,graphicx,amsmath,amssymb}
\renewcommand{\theequation}{\arabic{section}.\arabic{equation}}
\catcode`\@=11 \@addtoreset{equation}{section}\catcode`\@=12
\newcommand{\fnm}{\footnotemark}
\newcommand{\fnt}{\footnotetext}

\begin{document}
\LARGE
\begin{center}
{\bf \large On quasinormal modes in 4D black hole solutions in the model \\ with anisotropic fluid}

\vskip 0.4 cm
\small

S. V. Bolokhov\fnm[1]\fnt[1]{bol-rgs@yandex.ru}$^{,a}$
and  V. D. Ivashchuk\fnm[2]\fnt[2]{ivashchuk@mail.ru}$^{,a,b}$

\vspace{7pt}

 \it 

 (a) 
   Institute of Gravitation and Cosmology,
   Peoples' Friendship University of Russia
   (RUDN University), \\ 
   6 Miklukho-Maklaya Street,
   Moscow, 117198, Russian Federation \\
 (b) Center for Gravitation and Fundamental Metrology, \\
  VNIIMS, 46 Ozyornaya St., Moscow 119361, Russian Federation \\

\end{center}

%\vskip0.1cm
\normalsize

\begin{abstract}
We consider a family of 4-dimensional black hole  solutions from 
Dehnen et al. ( Grav. Cosmol. 9:153, arXiv: gr-qc/0211049, 2003)
governed by natural number $q= 1, 2, 3 , \dots$,   which appear in the model
with  anisotropic fluid  and the equations of state:  $p_r = -\rho (2q-1)^{-1}$, 
$p_t = - p_r$, where  $p_r$  and $p_t$  are  pressures in radial and transverse 
directions, respectively, and  $\rho > 0$ is the  density. These equations  of state obey weak, strong
and dominant energy conditions.    For $q = 1$ the metric of the solution coincides with that of   the Reissner-Nordstr\"om one.  The global structure  of solutions is outlined, 
giving rise to Carter-Penrose diagram of  Reissner-Nordstr\"om or Schwarzschild types for 
odd $q = 2k + 1$ or even $q = 2k$, respectively. Certain physical parameters corresponding to BH solutions 
(gravitational mass, PPN parameters, Hawking temperature and entropy)
are calculated.  We obtain and analyse  the quasinormal modes for a test massless scalar field in the eikonal approximation.  For limiting case $q = + \infty$, they coincide with the well-known results for the Schwarzschild solution.  We  show that the Hod conjecture which connect the Hawking temperature  and the damping rate  is obeyed for all $q \geq 2$ and all (allowed) values of parameters.

\end{abstract}

%\section{Using Other Packages}\label{aba:sec1}

%\bigskip

%%%%%%%%%%

\section{Introduction}
The decaying oscillations such as quasinormal modes (QNMs) \cite{4,5,6,BlMash,8,9,10,11,12,13,14} are at presence  a very interesting and popular topic of investigations. A possible application of QNMs may be related to   gravitational waves  \cite{1,2,3}  emitted during the ringdown (final) stage of binary black hole (BH) mergers. It is belived that the frequencies of gravitational waves may  be calculated 
by   using certain  superpositions of QNMs.  The importance of  these experiments is following: the  analysis  of experimental data may clarify the nature of gravity in the regime of strong fields. 

In solving the quasinormal  mode (QNM) problem for certain physical tasks (e.g. related to asymptotically flat black hole solutions) one should seek the solutions to a wave equation  of the form  $\Phi(t,x) = e^{-i\omega t} \Phi_{*} (x) $, where  $\Phi_{*}   = \Phi_{*}  (x)$ obeys a Schr\"odinger-type equation
\begin{equation}
\label{0.1}
\left(-\epsilon^2\frac{d^2}{dx^2} + V(x)\right) \Phi_{*}  = \omega^2 \Phi_{*},
\end{equation}
with  $ x \in  (- \infty, + \infty)$  usually appearing as tortoise coordinate  and $\epsilon > 0$, while typically $\epsilon = 1$  \cite{10,11,12,13}. 
 
For a certain class of  spherically symmetric solutions (which contain Schwarzschild,  Reissner-Nordstr\"om ones and the solutions considered in the body of the paper)
 the potential is a smooth function obeying  $V(x) >0$ , which  tends to  $0$ either  when $x \to - \infty$ (in approaching  to horizon) or $x \to +\infty$ (in approaching  to spatial infinity).  By choosing (typically) the QNM frequencies $\omega$ as  complex numbers obeying ${\rm Re} \ \omega > 0$ and ${\rm Im} \ \omega < 0$, one get  the wave functions $\Phi(t,x) = e^{-i\omega t} \Phi_{*} (x) $ to be damped in time as $t \to +\infty$,
while $|\Phi_{*}(x)|$ has an exponential growth (in $|x|$)  as $|x| \to \infty$.
The  QNMs  \cite{12,13} are usually calculated by a analytical  continuation method 
\cite{BlMash,8,9}. According to Ref. \cite{14}  this method reads as follows: 
one should  start with the Schr\"odinger equation  for a wave function $\Psi = \Psi(x)$ 
\begin{equation}
\label{0.2}
 \left(-\hbar^2\frac{d^2}{dx^2} - V(x)\right) \Psi = E \Psi.
\end{equation}
It  describes a (non-relativistic) quantum particle of mass $1/2$ ``moving'' in the potential $-V(x)$.
Let us suppose that the Schr\"odinger operator corresponding to (\ref{0.2}) has non-empty 
discrete spectrum  $E_n = E(\hbar, n| -V)$, where $n =  0, 1, \dots$.  
(The corresponding  eigen functions   $\Psi =  \Psi_n(x)$  should be  exponentially decaying  
as $ x \to \pm \infty $). Due to
 Ref. \cite{14} one should put for QNM frequencies
\begin{equation}
\label{0.3}
\omega^2 = -E(\hbar = i\epsilon, n| -V).
\end{equation}
Here $n = 0,1, \dots$ is called as overtone number.

In this article we deal with  4D black hole  solutions from Ref.  \cite{DIM}.
These solutions take place  in the model  with  anisotropic fluid 
with the following  equations of state:  
\begin{equation}
\label{0.4}
p_r = -\rho /(2q-1), \qquad   p_t =  \rho /(2q-1),
\end{equation}
where  $p_r$  and $p_t$  are  pressures in radial and transverse 
   directions, respectively,  $\rho > 0$ is the  density and 
$q= 1, 2, 3 , \dots$  is the natural number. (In (\ref{0.4}) we put $c =1$ .)  
It may be readily verified that 
these equations  of state obey weak ($\rho \geq 0$, $ \rho + p_i  \geq 0$), 
strong ($\rho + \sum_{j}  p_j  \geq 0$, $ \rho + p_i \geq 0$)
and dominant  ($\rho \geq |p_i | $)
energy conditions   (here $(p_i) = (p_r, p_t, p_t))$.    

Here we obtain and analyse  the QNMs for a test massless scalar field in the eikonal approximation
which is the main subject of the paper. By product we present the 
global structure  of BH solutions under consideration and 
calculate  certain physical parameters corresponding to them 
(gravitational mass, PPN parameters, Hawking temperature and entropy).

The paper is organised as follows. In Section 2  
we present the black hole  solutions from Ref.~\cite{DIM}. 
In Section 3 we analyse the global structure of the solutions.
In Section 4 we calculate certain physical parameters 
which correspond to the solutions under consideration.  
In Section 5 we find the  frequences of QNMs  in the eikonal approximation 
 which correspond  to massless test scalar field in the background metric 
 of our  BH solutions with anisotropic fluid for $q= 1, 2, 3 , \dots$.
Section 6 is devoted to special (integrable) cases $q = 1, 2, 3$ and the limiting case $q = + \infty$.
In Section 7 we verify the validity of the Hod conjecture \cite{Hod} 
for the solutions under consideration with $q >1$.

\section{The black hole solution}

Here we consider the solutions to Einstein equations
\begin{equation}
\label{1.0}
R^{\mu}_{\nu} - \frac{1}{2} \delta^{\mu}_{\nu} R = \kappa T^{\mu}_{\nu},
\end{equation}
 where $\kappa = 8 \pi G /c^4$, $G$ is Newton gravitational constant and $c$ is speed of light. 

The solutions under consideration are defined on the four-dimensional manifold with topology
\begin{equation}
\label{1.1}
  M = \mathbb{R}_{(\rm radial)}\times \mathbb{S}^{2}\times
\mathbb{R}_{(\rm time)}.
\end{equation}
Here the spherical coordinate system is used: $ x^\mu=(r, \theta, \phi, t)$  with signature $(+++,-)$.
The energy-momentum tensor of anisotropic fluid is taken as
\begin{equation}
\label{1.2}
 (T^{\mu}_{\nu})={\rm diag}\left(p_r,\ p_t, p_t, \ -\rho c^2 \right),
\end{equation}
and the equations of state read
\begin{equation}
\label{1.3}
  p_r = -\rho c^2 (2q-1)^{-1}, \qquad p_t = - p_r.
\end{equation}
Here $\rho$ is the mass density, $p_r$  and $p_t$  are  pressures in radial and orthogonal 
(to radial) directions, respectively.   

The parameter $q$ describes relations between the pressures and the mass density; $q>0$, $q\neq
1/2$. In the present paper, the parameter $q$ is taken to be a natural number to avoid the non-analytical
behaviour of the metric at the (would be) horizon.

The solution has the following form \cite{DIM}:

\begin{equation} 
\label{1.4}
  ds^2= g_{\mu \nu} dx^{\mu} dx^{\nu}  = (H(r))^{2/q}\left[
 \frac{dr^2}{1-\frac{2\mu}{r}}
 + r^2 d\Omega^2
 -(H(r))^{-4/q}\left(1-\frac{2\mu}{r}\right) c^2 dt^2\right],
\end{equation}
\begin{equation}
 \label{1.5}
 \kappa \rho c^2 = \frac{(2q-1)P(P+2\mu)(1-2\mu r^{-1})^{q-1}}{H(r)^{2+\frac{2}{q}}\; r^{4}},
\end{equation}
where the function $H(r)$ reads as follows:
\begin{equation}
 \label{1.6}
 H(r) = 1+\frac{P}{2\mu}
 \left[1-\left(1-\frac{2\mu}{r}\right)^q \right].
\end{equation}
The metric on the sphere $\mathbb{S}^2$ is denoted by
 $d\Omega^2$; parameters  $P,\mu > 0$ are arbitrary. Originally we put $r > 2\mu = r_h$
 but the domain of definition of the metric will be extended below. 

The equations of motion (\ref{1.0}) imply the following relation for the
scalar curvature
\begin{equation}
\label{1.7}
R[g] = - \kappa T^{\mu}_{\mu} = \frac{2(q-1)}{2 q -1} \kappa \rho c^2,
\end{equation}
which will be used below for identifying the singularities of solutions
for $q = 2,3,4,\dots$.

\section{The global structure of the solution}

%\subsection{An apparent extension of the solution}

In what follows we will use the following relation for the metric
\begin{equation} 
\label{1.4s}
  ds^2=  - A(r) c^2 dt^2  + (A(r))^{- 1} dr^2 + C(r) d\Omega^2 ,
\end{equation}
where
\begin{eqnarray} 
\label{1.4a}
 A = A(r) = (H^2(r))^{- 1/q} \left(1-\frac{2\mu}{r}\right), \\
 C = C(r) = (H^2(r))^{1/q} r^2 . \label{1.4b}
\end{eqnarray}
Here $A = A(r)$ is so-called ``red shift function'', $C(r) > 0$ is ``area function''.

%\subsection{Carter-Penrose diagrams}

The global structure of the solutions
above may be studied by analysing the behaviour of the
``redshift function'' ($A(r)$) and the ``area function'' $C(r)$ (the
factor at $d\Omega^2$) at critical points corresponding to
horizons or singularities.
The Carter-Penrose diagrams can be constructed for various values 
of the parameters using the standard algorithm  \cite{BR}.
For our BH solutions it was done in Ref. \cite{IvasBol-Global}.

 In what follows we denote by $r=r^\star$ the maximal root of the equation
$H(r)=0$. We have $r^{\star} < 0$ for odd $q = 2k + 1$ and 
$r^{\star}  > 0$ for  even $q = 2k$.

There are three classes of important critical points of the radial
coordinate $r$ for the metric (\ref{1.4s}):

1) $r=r_h \equiv{2\mu}$. This point corresponds to a
regular external horizon.

2) $r=r^{\star}$. This point corresponds to the singularity.

3) $r=0$ for odd $q = 2k + 1$. This point corresponds to internal horizon.

We introduce the following notations. Let ${\bf Sch}[r_1,r_2]$
($r_1<r_2$) be a Carter-Penrose diagram of Schwarzschild type with a singularity
at a point $r_1$ and a regular horizon at $r_2$. (Fig. 1.) Similarly, we
denote by ${\bf RN}[r_1, r_2, r_3]$ ($r_1<r_2<r_3$) the diagram of
Reissner-Nordstr\"om type with  singularity at $r_1$,  an internal
horizon at $r_2$, and an external horizon at $r_3$. (Fig. 2.)

As a result of analysis, we conclude that the structure of
diagrams depends mostly on the parity of the parameter $q$:

\begin{itemize}
\item For $q=2m,\,m\in\mathbb{N}$, we have a diagram of type ${\bf
Sch}[r^\star,r_h]$.
\item For $q=2m+1$ the diagram is of type ${\bf RN}[r^\star, 0, r_h]$. 
\end{itemize}
\begin{figure}[h]
\center
\includegraphics[width=0.4\linewidth]{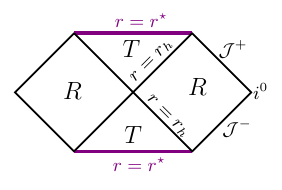}
\caption{Carter-Penrose diagram for case $q=2,4,6,...$}
\end{figure}
\begin{figure}[h]
\center
\includegraphics[width=0.3\linewidth]{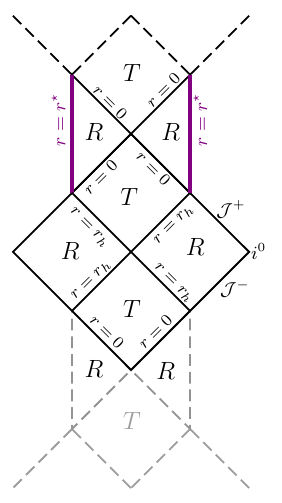}
\caption{Carter-Penrose diagram for case $q=1,3,5,...$}
\end{figure}

{\bf Extremal case.} 
Let us consider an extremal case of the solution under consideration when $\mu \to + 0$.
By using relations (\ref{1.4}), (\ref{1.5}) and (\ref{1.6}) we get in the limit $\mu \to + 0$

\begin{equation} 
\label{1.4e}
  ds^2= g_{\mu \nu} dx^{\mu} dx^{\nu}  = (H_{e}(r))^{2/q}\left[  dr^2  + r^2 d\Omega^2
         -(H_{e}(r))^{-4/q} c^2 dt^2\right],
\end{equation}
\begin{equation}
 \label{1.5e}
 \kappa \rho c^2 = \frac{(2q-1)P^2}{(H_{e}(r))^{2+\frac{2}{q}}\; r^{4}},
\end{equation}
where 
\begin{equation}
 \label{1.6e}
 H_{e}(r) = 1+\frac{P q}{r},
 \end{equation}
with $P > 0$. For $q > 1$ the metric (\ref{1.4e}) describes a naked singularity 
corresponding to $r \to +0$. Indeed, using relations (\ref{1.7}) and
(\ref{1.5e}) we obtain for the scalar curvature
\begin{equation}
\label{1.7e}
R[g] =  \frac{2(q-1)P^2}{(r + Pq)^{2+\frac{2}{q}}\; r^{2 - \frac{2}{q}}}.
\end{equation}
For $q = 2,3,4,\dots$ we are led to relation: $R[g] \to + \infty$ 
as $r \to +0$, which tells us about the singularity corresponding to $r = +0$.
For $q =1$ the metric (\ref{1.4e}) is coinciding with the metric of extremal
Reissner-Nordstr\"om solution with ``double'' horizon corresponding to $r = +0$
and singularity (center) at $r = -P +0$.

\section{Physical parameters}

In this section we deal with some physical parameters of the solutions.
Here we put for simplicity $c = \hbar = k_B = 1$.

\subsection{Gravitational  mass and PPN parameters}

 Let us
consider the 4-dimensional space-time with the metric
(\ref{1.4}) for $r > 2 \mu$. Introducing a new radial variable $\bar{R}$ by the relation:

\begin{equation} 
 \label{3.7} 
r = \bar{R} \left(1 + \frac{\mu}{2\bar{R}} \right)^2, 
\end{equation}

we rewrite the metric in the following form:

\begin{eqnarray} 
 \label{3.8}
 ds^2 = H^{2/q} \left[ - H^{- 4/q }
 \left(\frac{1-\frac{\mu}{2 \bar{R}}}{1+\frac{\mu}{2 \bar{R}}}\right)^2 dt^2
 + \left( 1+\frac{\mu}{2 \bar{R}} \right)^4 \delta_{ij}
 dx^i dx^j \right]
\end{eqnarray}
$i,j = 1,2,3$. Here $\bar{R}^2 = \delta_{ij} x^i x^j$.

The parametrized post-Newtonian (Eddington) parameters are
defined by the well-known relations

\begin{eqnarray} 
 \label{3.9}
 g_{00} = - (1- 2V + 2\beta V^2) + O (V^3), \\
 \label{3.10} g_{ij} = \delta_{ij} (1 + 2 \gamma V) + O(V^2),
\end{eqnarray}
 $i,j = 1,2,3$. 
 Here
\begin{equation} 
 \label{3.11}
  V= \frac{GM}{\bar{R}}
\end{equation}
is the Newtonian potential, $M$ is the gravitational mass and $G$ is
the gravitational constant.

From (\ref{3.8})-(\ref{3.10}) we obtain:
\begin{equation} 
 \label{3.12}
 GM = \mu +  \frac{P}{q} 
\end{equation}
and
\begin{eqnarray} 
 \label{3.13}
 \beta - 1= \frac{q A_f }{2(GM)^2}, \\
 \label{3.13a}
 \gamma = 1,
\end{eqnarray}
where  
\begin{equation}
A_f =  P (P + 2 \mu),  \label{3.13A}
\end{equation}
or, equivalently, 
$P = - \mu + \sqrt{\mu^2 + A_f} > 0$. 

The parameter $\beta -1$ is proportional to the  ratio of two physical
parameters: the anisotropic fluid density parameter $A_f$
and the gravitational radius squared $(GM)^2$.

\subsection{Hawking temperature and entropy}

The Hawking temperature of the black hole may be calculated
using the well-known relation \cite{York}
\begin{equation} 
 \label{3.Y}
      T_H  = \frac{1}{4 \pi \sqrt{-g_{00} g_{rr}}}
      \frac{d(-g_{00})}{dr} \Biggl|_{_{\ horizon}},  
 \end{equation} 
where here $g_{rr} = (A(r))^{- 1}$, see (\ref{1.4s}).      

We get
\begin{equation} 
 \label{3.H}
  T_H = \frac{1}{8 \pi \mu}   \left(1 + \frac{P}{2\mu} \right)^{- 2/q}.
\end{equation} 
Here $q = 1,2, \ldots.$

The Bekenstein-Hawking (area) entropy $S = {\cal A}/(4G)$, corresponding to
the horizon at $r = 2 \mu$, where ${\cal A}$ is the horizon area, reads
\begin{equation} 
 \label{3.S}
  S_{BH} = \frac{4 \pi \mu^2}{G}   \left(1 + \frac{P}{2\mu} \right)^{2/q} .
\end{equation}

%%%%%%%%%%%%%%%%%%%n%%%%%%%%%%%%%%%%%%%%%%%%%%%%%%%%%%%%%%%%%%%%%
\section{Quasinormal modes}\label{qnms}
%%%%%%%%%%%%%%%%%%%%%%%%%%%%%%%%%%%%%%%%%%

In this section we derive quasinormal modes (in eikonal approximation) for our static
and spherically symmetric solution (for given $q$) with the metric given (initially) 
in the following general form

\begin{equation}\label{4.2m}
 ds^2 =-A(u) dt ^2 + B(u) du ^2 + C(u) d \Omega ^2 \ ,
\end{equation}
where $A(u),$ $B(u),$ $C(u)>0$  and $d \Omega^2 = d \theta^2 + \sin^2 \theta d \phi^2$.
Note that in this section and below we  use the Planck units, i.e. we put
$\hbar = G=c=1$.

We consider  a test massless scalar field defined in the background given
by the metric \eqref{3.8}. The equation of motion in general is written in the form of the
covariant Klein-Fock-Gordon equation
\begin{equation}\label{4.2}
\Delta \Psi =\frac{1}{\sqrt{\vert g \vert }}\partial_{\mu
}(\sqrt{\vert g\vert }g^{\mu \nu } \partial_{\nu } \Psi )=0.
\end{equation}
where $\mu, \nu=0, 1, 2, 3$.
In order to solve this equation we separate variables in function $\Psi$ as follows
\begin{equation}\label{4.3}
\Psi = e^{-i \omega t}e^{-\gamma}\Psi_{*}(u)Y_{lm},
\end{equation}
where $Y_{lm}$ are the spherical harmonics, $l$ is the multipole quantum number, $l = 0,1, \dots $
and $m = - l, \dots, 0, \dots, l$. 

Equation \eqref{4.2}, after using \eqref{4.3} yields
the equation describing the radial function $\Psi_{*}(u)$ and having a Schr\"odinger-like form
\begin{equation}\label{4.4}
\frac{d^2 \Psi_*(u)}{du^2}+\bigg\{\frac{B}{A}\omega^{2}
-\frac{B}{C}l(l+1)-\gamma''-(\gamma ')^{2}\bigg\} \Psi_*(u)=0
\end{equation}
where
\begin{equation} \label{4.G}
\gamma =\frac{1}{2}\ln (B^{-1}C\sqrt{AB})
\end{equation}
and  $\gamma'=d\gamma/du$, $\gamma^{''}=\frac{d^{2}\gamma }{du^{2}}$.

Taking into account above expressions one can examine our black hole solution
which has the following form
\begin{equation} \label{4.ds}
 ds^2 =-A(r) dt ^2 + \frac{dr^2}{A(r)} + H^{a} r^2 d \Omega ^2 ,
\end{equation}
where $A(r)$ and $C(r)$ according to Eq. (\ref{1.4}) can be written as
\begin{eqnarray} \label{4.f}
 A(r)&=&A=H^{-a}\left( 1-\frac{2\mu}{r} \right), \\
 C(r)&=&C=H^{a} r^2 = \exp(2 \gamma), \qquad a = 2/q,    \label{4.C}
\end{eqnarray}
where  
 \begin{equation}
 \label{4.H}
 H(r) = 1+\frac{P}{2\mu} \left[ 1-\left(1-\frac{2\mu}{r}\right)^q \right] = 1 + p (1 - z^q) 
  \end{equation}
is the moduli function, $\mu  > 0$, $P > 0$, $p = P/(2 \mu)$, $q = 1,2, \dots$ and 
\begin{equation}
 \label{4.Hz}
  z = 1-\frac{2\mu}{r},  \qquad  r = \frac{2\mu}{1- z}.
  \end{equation}
We note that $0 < z < 1$ for $r > 2\mu$. 

After using the ``tortoise'' coordinate transformation
\begin{equation}
dr_*= \frac{dr}{A(r)}
\end{equation}
the metric takes the following form
\begin{equation}
  ds^2 =- A dt ^2 + A dr_*^2 + C d \Omega ^2 \ . 
 \end{equation}
For the choice   of the tortoise coordinate as a radial one ($u =r_*$) we have $A=B$ and
\begin{equation}
\gamma =\frac{1}{2}\ln C = \frac{1}{2}\ln (H^{a} r^2),
\end{equation}
$a = 2/q$.

Thus, the Klein-Fock-Gordon equation becomes
\begin{equation}\label{4.5}
 \frac{d^{2}\Psi_{*}}{dr_{*}^{2}}+\big\{ \omega ^{2}-V \big\} \Psi_{*}=0,
\end{equation}
where $\omega$ is the (cyclic) frequency of the quasinormal mode and
$V =V(r) = V(r(r_{*}))$ is the effective potential
\begin{eqnarray} \label{4.5V}
V&=&\mathcal{V}+\delta\mathcal{V},
 \\ \mathcal{V} &=&\frac{l(l+1)A}{H^{a}r^2} \label{4.5Ve}
 \\  &=& \frac{l(l+1) z (1-z)^2(1 + p (1-z^q))^{-2a}}{(2\mu)^2}, \nonumber
 \\ \delta\mathcal{V} &=& \gamma''+(\gamma ')^{2} = (\sqrt{C})^{''} / \sqrt{C},
  \label{4.5Vd}
\end{eqnarray}
so that $\mathcal{V}$ is the eikonal part of the effective potential.
Here and below we denote $F' = \frac{dF}{dr_{*}}= A \frac{dF}{dr}$.

In what follows we consider the so-called eikonal approximation  when $l \gg 1$.

The maximum of the eikonal part of the effective potential is found from the extremum condition
\begin{eqnarray} \label{4.6V}
\mathcal{V}'=  A \frac{d \mathcal{V}}{dr} = 0
\end{eqnarray}
or, equivalently,
\begin{equation} \label{4.6VV}
\nu = \frac{d \mathcal{V}}{\mathcal{V} dz}  
 = \frac{1}{z} - \frac{2}{1 -z}  + \frac{4 p z^{q-1} }{1 + p (1-z^q)}  =0,
\end{equation}
or 
\begin{equation} \label{4.6m}
  p z^{q+ 1} - 3 p z^{q} + (1+ p)(3z - 1) = 0.
  \end{equation}

{ \bf Proposition 1}. {\em For any $P> 0$, $\mu > 0$ and $q \in {\bf N}$,  the 
extremality relation (\ref{4.6V}) has only one solution
for $r > 2 \mu$, which is the point of maximum for $\mathcal{V}(r)$. 
}

The proposition  is proved in Appendix. We denote this 
point of extremum by $r_0$.  In terms of variable $z$ we get
that the point  $z_0 = 1- 2\mu/r_0$ is a unique solution
to Eq.  (\ref{4.6m}) for  $z \in (0,1)$.

The maximum of the eikonal part  of the effective potential thus becomes
\begin{equation}
   \mathcal{V}_0=\mathcal{V}(r_0) = \frac{l(l+1)}{H^{2a}(r_0) r_0^2}
    \left(1-\frac{2\mu}{r_0}\right).   \label{4.9V}
\end{equation}

\begin{figure}[h!]
	\begin{center}
		\includegraphics[width=1\linewidth]{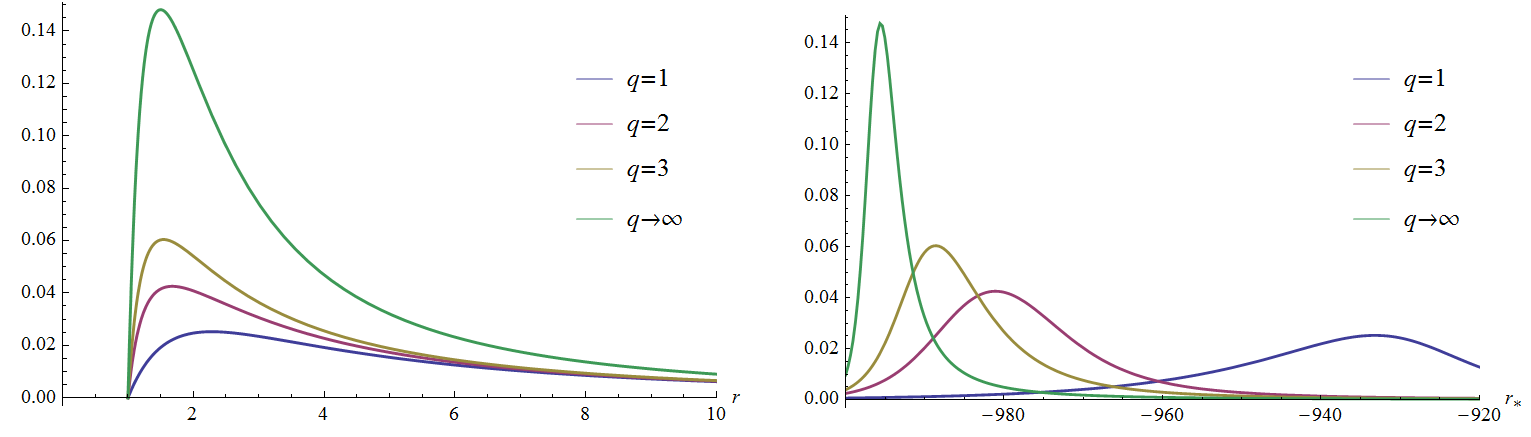}
		\caption{The graphical representation of the reduced potential $\mathcal{V}/(l(l+1))$ as a function of the radial coordinate $r$ (left panel) and the tortoise coordinate $r_*$ (right panel) for $P=2\mu=1$, $q = 1,2,3$ and the limiting case $q\to+\infty$.}
		\label{Vplot}
	\end{center}
 \end{figure}

In Fig.~\ref{Vplot} we plot the reduced eikonal part of the effective potential $\mathcal{V}/(l(l+1))$ ($l \neq 0$)
as a function of the radial coordinate $r$ (left panel) and the tortoise coordinate $r_*$ (right panel).

As can be seen from examples presented in figure for special fixed values of $P$ and $\mu$, 
 the maximum of the effective potential is largest for $q=+\infty$ case and smallest for $q = 1$ case.
 The case with $q=2$ is in the middle. At large distances the effective potential tends to zero, as expected.

The second derivative with respect to the tortoise coordinate in the point of extremum is given by
\begin{eqnarray}
\mathcal{V}_0''&=&\frac{d^2\mathcal{V}}{dr_*^2}\bigg|_{r_{*} =r_{*}(r_0)}
  = A_0^2 \frac{d^2\mathcal{V}}{dr^2}\bigg|_{r=r_0}
   \nonumber \\
 &=&  A_0^2 \left(\frac{dz}{dr}\right)^2 \bigg|_{r=r_0} \frac{d^2\mathcal{V}}{dz^2}\bigg|_{z=z_0}
 =  A_0^2 \left(\frac{2\mu}{r_0^2}\right)^2 \frac{d^2\mathcal{V}}{dz^2}\bigg|_{z=z_0}  \label{4.V2a}
\end{eqnarray}
where $A_0 = A(r_0)$, see (\ref{4.f}). The calculation of second derivative
gives us
\begin{equation}
\frac{d^2\mathcal{V}}{dz^2}\bigg|_{z=z_0} = \frac{d}{dz} \left( \nu \mathcal{V} \right)\bigg|_{z=z_0}
= \mathcal{V}_0 \frac{d \nu }{dz} \bigg|_{z=z_0}, \label{4.V2b}
\end{equation}
where $\nu = \nu(z)$ is defined in (\ref{4.6VV}). We get 
 \begin{eqnarray}
   \frac{d \nu }{dz} \bigg|_{z=z_0} =  
   - \frac{1}{z_0^2} - \frac{2}{(1 -z_0)^2} \nonumber 
   \\  + \frac{4 p^2 q z_0^{2q-2} }{(1 + p (1-z_0^q))^2} 
   +  \frac{4 p (q -1) z_0^{q-2} }{1 + p (1-z_0^q)}. 
         \label{4.n1}
  \end{eqnarray}
The last two terms in this relation may be simplified by using the relation for the third 
term in (\ref{4.6VV}). We obtain
\begin{eqnarray}
   \frac{d \nu }{dz} \bigg|_{z=z_0} =  
   - \frac{1}{z_0^2} - \frac{2}{(1 -z_0)^2}  \nonumber  \\ 
    + \frac{q}{4} \left( \frac{1}{z_0} - \frac{2}{1 -z_0}\right)^2
       +   \frac{q-1}{z_0} \left( -\frac{1}{z_0} + \frac{2}{1 -z_0} \right)
      \nonumber     \\
      = - \frac{3q}{4 z_0^2} + \frac{q - 2}{z_0 (1 -z_0)^2}.  
               \label{4.n3}
  \end{eqnarray}
 
Thus, by using  (\ref{4.V2a}), (\ref{4.V2b}) and  (\ref{4.n3})
we find
\begin{equation}
\mathcal{V}_0''  =  - \frac{1}{2} A_0^2 \left(\frac{2\mu}{r_0^2}\right)^2 \mathcal{V}_0
                  \mathcal{B}(z_0), 
 \label{4.V2}
\end{equation}
 where
\begin{equation}
  \mathcal{B}(z) = \frac{3}{2} q - \frac{2 (q - 2 ) z}{(1-z)^2}. \label{4.B}
\end{equation}

The square of the cyclic frequency in the eikonal approximation reads as following \cite{12,13}
\begin{equation}
\omega^{2}=\mathcal{V}_0-i \left(n+\frac{1}{2}\right) \sqrt{-2 \mathcal{V}_0''} + O(1),
\label{4.om}
\end{equation}
where $l \gg 1$ and $l \gg n$. Here $n = 0, 1, \dots$ is the overtone number. By choosing an appropriate
sign for  $\omega$ we get the asymptotic relations (as $l \to + \infty$)
on real and imaginary parts of complex  $\omega$ in the
eikonal approximation
\begin{eqnarray}
  {\rm Re}(\omega)  &=&   \left(l + \frac{1}{2}\right) H_{0}^{-a} r_0^{-1} z_0^{1/2} 
      + O\left(\frac{1}{l+\frac{1}{2}}\right),  \label{4.11Re}
      \\
 {\rm Im}(\omega)  &=&  - \left(n + \frac{1}{2}\right) 
 H_{0}^{-a} \mu r_0^{-2} \mathcal{B}_0^{1/2}  
    + O\left(\frac{1}{l+\frac{1}{2}}\right), \label{4.11Im}
\end{eqnarray}
where $H_0 = H (r_0)$ (see  (\ref{4.H})),  $r_0 =  2\mu /(1 - z_0)$, and  $z_0 \in (0,1)$ is solution 
to master equation (\ref{4.6m}) and $\mathcal{B}_0 = \mathcal{B}(z_0)$, where 
$\mathcal{B}(z)$ is defined in (\ref{4.B}).

We note that  the parameters of the unstable circular null geodesics
around  stationary spherically symmetric and asymptotically flat black holes are 
in correspondence with the eikonal part of  quasinormal modes of these black holes.
See \cite{40,41,VE} and references therein.
Due to Ref. \cite{42} this correspondence is valid if certain restrictions on perturbations are imposed.

\section{Special cases $q = 1, 2, 3$ and the limiting case $q = + \infty$}

In this section we consider eikonal QNM for three cases $q = 1, 2, 3$ when the master equation 
(\ref{4.6m}) may be solved in radicals for all values of $p > 0$  
and also in the limiting case $q = + \infty$.

\subsection{The case $q = 1$}
 Let us consider the case  $q = 1$ ($a = 2)$.
 In this case the master equation (\ref{4.6m}) is just a quadratic one 
 with two roots: 
 \begin{equation} \label{4.6zpm}
   z_{+} = \frac{-3 \pm \sqrt{4 p(p+1)+9}}{2p},
   \end{equation}
  Here 
  \begin{equation} \label{4.6z0}
    z_{+} = z_0 = z_0(1,p),
    \end{equation}
 is belonging to interval $(0,1)$, while $z_{-} < 0$ is irrelevant for our consideration. 
 We have 
 \begin{equation} \label{4.6derz0}
    \frac{\partial z_0(1,p)}{\partial p} 
    = \frac{3\sqrt{4 p(p+1) + 9} - 2p - 9}{2p^2 \sqrt{4 p(p+1) + 9} } > 0.
    \end{equation}
 We get that the fuction  $z_0(1,p)$ is monotonically increasing and have the following limits:
 $z_0(1,p) \to 1/3$ as $p \to +0$ and $z_0(1,p) \to 1$ as $p \to + \infty$.
 For all values $p > 0$ we have 
  \begin{equation} \label{4.6bz0}
     1/3 < z_{0}(1,p) < 1.
  \end{equation}

In this case the eikonal QNM (see (\ref{4.11Re}) and (\ref{4.11Im})) read 
\begin{eqnarray}
  {\rm Re}(\omega)  &=&   \left(l + \frac{1}{2}\right) H_{0}^{-2} r_0^{-1} z_0^{1/2} 
      + O\left(\frac{1}{l+\frac{1}{2}}\right),  \label{4.11Req1}
      \\
 {\rm Im}(\omega)  &=&  - \left(n + \frac{1}{2}\right) 
 H_{0}^{-2} \mu r_0^{-2} \sqrt{\frac{3}{2} + \frac{2 z_0}{(1-z_0)^2}} 
 %\nonumber 
    + O\left(\frac{1}{l+\frac{1}{2}}\right), \label{4.11Imq1}
\end{eqnarray}
where $H_0 =  1+\frac{P}{r_0}$ , 
 $r_0 =  2\mu /(1 - z_0)$ and $z_0 = z_0(1,p)$ is defined in (\ref{4.6z0}). 

It may be verified that relations (\ref{4.11Req1}), (\ref{4.11Imq1})
may be rewritten as follows

\begin{eqnarray}
{\rm Re}(\omega)&=&\left(l+\frac{1}{2}\right)\sqrt{\frac{\bar{M}}{\bar{r}_0^3}-\frac{Q^2}{2 \bar{r}_0^4}}+O\left(\frac{1}{l+\frac{1}{2}}\right), \label{4.12Req1}\\
{\rm Im}(\omega)&=&-\left(n+\frac{1}{2}\right)\sqrt{\frac{\bar{M}}{\bar{r}_0^3}
-\frac{Q^2}{2 \bar{r}_0^4}} \sqrt{\frac{3 \bar{M}}{\bar{r}_0}-\frac{2 Q^2}{\bar{r}_0^2}}\nonumber
\\&+&O\left(\frac{1}{l+\frac{1}{2}}\right) \label{4.12Imq1}, 
\end{eqnarray}
where $\bar{r}_0= r_0 + P$, $\bar{M} = \mu + P = G M$ and 
\begin{equation}
 A_f = P (P + 2 \mu) = \frac{1}{2} Q^2     \label{4.12Q}.  
\end{equation}

Here $\bar{r}_0$ corresponds to the position of the unstable, 
circular photon orbit in the Reissner-Nordstr\"om spacetime with the metric
  \begin{equation} 
   ds^2 =   - \bar{f}(\bar{r}) dt^2 +   (\bar{f}(\bar{r}))^{-1} d\bar{r}^2 + \bar{r}^2  d \Omega^2_{2},
           \label{4.12RN}
  \end{equation}
 where $\bar{f}(\bar{r}) = 1 - \frac{2GM}{\bar{r}} + \frac{Q^2}{2 \bar{r}^2}$, 
 with $Q^2$ given by  (\ref{4.12Q}). Our AF (anisotropic fluid) metric (\ref{1.4}) 
 for $q = 1$ is coinciding with the Reissner-Nordstr\"om one  (\ref{4.12RN})
 when the following relation for radial coordinates $\bar{r}= r + P$ is imposed. 
 
Relations (\ref{4.12Req1}), (\ref{4.12Imq1}) for Reissner-Nordstr\"om spacetime 
were obtained in Ref.~ \cite{AndOn} for $n =0$. 

\subsection{The case $q = 2$}
 
 Now we put $q=2$ ($a = 1$). The master equation (\ref{4.6m})
 in this case is just cubic one. It has a unique (real) solution 
 $z_0 = z_0(2,p)$ for any $p >0$ belonging to interval $(1/3,1)$
  \begin{equation} \label{4.13z0}
    z_0 = z_0(2,p) = Z^{1/3} - p^{-1} Z^{-1/3} +1,
  \end{equation}
where
 \begin{equation} \label{4.13Z}
     Z = Z(p) = \frac{1}{p} \left( \sqrt{ 1 + \frac{1}{p}} - 1 \right).
   \end{equation}

The function $Z(p)$ is monotonically decreasing from $+ \infty$ to $+0$ and  has  the 
asymptotics: \\
$Z(p) \sim p^{-3/2} (1 - \sqrt{p} + O(p))$ as $p \to +0$
and  $Z(p) \sim 2^{-1} p^{-2} $ as $p \to + \infty$ which imply
$z_0(2,p) \to 1/3$ as $p \to +0$ and $z_0(2,p) \to 1$ as $p \to + \infty$.
It may be verified that the function $z_0(2,p)$ 
is monotonically inreasing from $1/3$ to $1$. 

The eikonal QNM for $q =2$ read 
\begin{eqnarray}
  {\rm Re}(\omega)  &=&   \left(l + \frac{1}{2}\right) H_{0}^{-1} r_0^{-1} z_0^{1/2} 
      + O\left(\frac{1}{l+\frac{1}{2}}\right),  \label{4.13Req1}
      \\
 {\rm Im}(\omega)  &=&  - \left(n + \frac{1}{2}\right)  H_{0}^{-1} \mu r_0^{-2} \sqrt{3} 
 %\nonumber 
    + O\left(\frac{1}{l+\frac{1}{2}}\right), \label{4.13Imq2}
\end{eqnarray}
where $H_0 =  1+\frac{P}{2\mu} \left[ 1-\left(1-\frac{2\mu}{r_0}\right)^2 \right]$ , 
 $r_0 =  2\mu /(1 - z_0)$, and  $z_0 = z_0(2,p)$ is defined in (\ref{4.13z0}). 

\subsection{The case $q = 3$}

Let us consider the last case $q=3$, when the master equation (\ref{4.6m})
of fourth power has a solution in radicals (which was obtained by Mathematica):
 \begin{eqnarray} 
       z_0 = z_0(3,p) = \frac{1}{2} \sqrt{X} - \frac{\sqrt{Y}}{4\sqrt{3}} + \frac{3}{4}, 
       \label{4.14z0} \\
     Y = 12 Z^{1/3} + 27 + 20 \left(1 + \frac{1}{p} \right) Z^{- 1/3},  \label{4.14Y}  \\ 
     Z = \frac{p+1}{2 p^2} ( 9 + 3^{-3/2} \sqrt{2187 - 500 p (p+1)}  ), \label{4.14Z} \\
     X = - \frac{3 \sqrt{3}}{2} \left(1 -  \frac{8}{p}\right)  Y^{-1/2} - Z^{1/3} \nonumber \\ 
         -  \frac{5(p+1)}{3 p} Z^{- 1/3} +  \frac{9}{2}.    \label{4.14X} 
  \end{eqnarray}

\begin{figure}[!h]
	\begin{center}
		\includegraphics[width=0.6\linewidth]{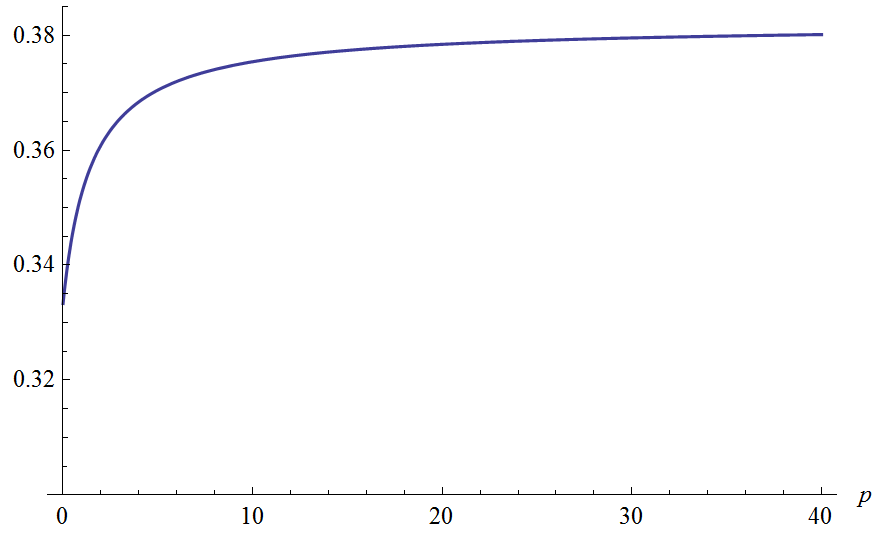}
		\caption{The graphical representation of the function  $z_0=z_0(3,p)$.}
		\label{z0}
	\end{center}
 \end{figure}

It may be verified that $z_0 = z_0(3,p)$, given by relations (\ref{4.14z0})-(\ref{4.14X}), 
is real for all $p > 0$ and obey $1/3 < z_0 < \frac{3 - \sqrt{5}}{2} \approx 0,382$. This property is graphically illustrated on Fig.~\ref{z0}. 

Relations (\ref{4.11Req1}), (\ref{4.11Imq1}) in this case reads as follows
\begin{eqnarray}
  {\rm Re}(\omega)  &=&   \left(l + \frac{1}{2}\right) H_{0}^{-2/3} r_0^{-1} z_0^{1/2} 
      + O\left(\frac{1}{l+\frac{1}{2}}\right),  \label{4.11Req3}
      \\
 {\rm Im}(\omega)  &=&  - \left(n + \frac{1}{2}\right) 
 H_{0}^{-2/3} \mu r_0^{-2} \left(\frac{9}{2} - \frac{2 z_0}{(1-z_0)^2}\right)^{1/2} 
 %\nonumber 
    + O\left(\frac{1}{l+\frac{1}{2}}\right), \label{4.11Imq3}
\end{eqnarray}
where $H_0 =  1+\frac{P}{2\mu} \left[ 1-\left(1-\frac{2\mu}{r_0}\right)^3 \right]$ , 
 $r_0 =  2\mu /(1 - z_0)$, and  $z_0 = z_0(3,p)$ is given by  (\ref{4.14z0}).

\subsection{The case $q = + \infty$}

In this case  the relations (\ref{4.11Re}) and (\ref{4.11Im}) for QNM in eikonal approximation read as follows

\begin{eqnarray} \label{4.15Req1}
 {\rm Re}(\omega)&=&\left(l+\frac{1}{2}\right)\sqrt{\frac{\mu}{r_0^3}}+O\left(\frac{1}{l+\frac{1}{2}}\right),\\
{\rm Im}(\omega)&=&-\left(n+\frac{1}{2}\right){\sqrt{\frac{\mu}{r_0^3}}}+O\left(\frac{1}{l+\frac{1}{2}}\right),
 \label{4.15Imq1}
\end{eqnarray}
where $r_0= 3 \mu = 3GM$ corresponds  the position 
where the black hole effective potential attains its maximum. 
We note that $r_0=3 \mu$  is the radius of the photon sphere 
for  the  Schwarzschild black hole with the metric   
 \begin{equation} 
   ds^2 =   - \left(1 - \frac{2 \mu}{r}\right) dt^2 +   \left(1 - \frac{2 \mu}{r}\right)^{-1} dr^2 + 
   r^2  d \Omega^2_{2},
           \label{4.12Sch}
  \end{equation}
which is coinciding with limiting case of our AF  metric (\ref{1.4}) when $q = + \infty$.

We note that relations (\ref{4.15Req1}), (\ref{4.15Imq1}) for Schwarzschild spacetime 
were obtained in Ref.~ \cite{BlMash}.

{\bf Remark.}  Here we restrict our choice  of a test field  by a massless (spin-zero, non-charged) scalar field which is the simplest ``perturbation'' to study. It may be shown that the consideration of a test Maxwell field on our black hole background will lead us to   two equations on  functions: $\Psi_{*,a} = a_{lm}( r_{*})$ and 
$\Psi_{*,b}  = b_{lm}( r_{*})$, which are certain combinations of coefficients (and their derivatives) coming from decomposing of  vector potential  in (vector) spherical harmonics. These equations (one of them is just an integrability condition) look like eq. (\ref{4.5}) but with another potential $V = {\cal V}$, instead of (\ref{4.5V})($\delta V = 0$ in this case). Thus, we will obtain the same spectrum  of  QNM in  eikonal approximation for a  test Maxwell field as for a massless  scalar field  considered here.

\section{Hod conjecture}\label{Hod}

Here we verify the conjecture by Hod \cite{Hod} on the existence of quasi-normal modes
obeying the inequality 
\begin{equation} \label{5.H}
  |{\rm Im}(\omega)  | \leq \pi T_H,
\end{equation}
where $T_H$ is Hawking temperature.

We note  the Hod conjecture has been tested  in theories with higher curvature corrections such as the Einstein-Dilaton-Gauss-Bonnet and Einstein-Weyl for the Dirac  field (with positive result) \cite{Zinhailo2019}.
(For negative result see Ref. \cite{45}.) 
Recently, we have also verified the Hod conjecture (with positive result) for a solution with 
dyon-like dilatonic black hole \cite{MBI} for certain values of dimensionless parameter $a \in [0,1]$.   
 
Here we verify this conjecture  by using the obtained eikonal relations  (\ref{4.11Im}) 
for  ${\rm Im}(\omega) $
and the relation for the Hawking temperature (\ref{3.H}). For our purpose it is sufficient
to check the validity of the inequality 
\begin{eqnarray} 
 y = y(p,q)\equiv \frac{|{\rm Im}(\omega_{\rm eik})(n=0) |}{\pi T_H} = \left[\frac{1+ p}{1+ p(1-z_0^q)}\right]^{2/q} \times 
  \nonumber \\ 
 \times  (1 - z_0)^{2} \sqrt{\frac{3}{2} q - \frac{2 (q - 2 ) z_0}{(1-z_0)^2}} < 1, \label{5.1}
\end{eqnarray}
for all $p = P/\mu > 0$, $q = 2,3,\dots$,  where  $z_0 = z_0(p,q)$ is unique solution to master 
equation (\ref{4.6m}), which obeys $0 < z_0 < 1$, see Lemma in Appendix. 

In (\ref{5.1}) we use the limiting  ``eikonal value'' given by the first term in (\ref{4.11Im})  for
the lowest overtone number $n=0$. 

{\bf Proposition 2.} The dimensionless parameter $y = y(p,a)$ from (\ref{5.1})  obeys the inequality: $y < 1$ for all $p > 0$ and $q \in \{2,3,4, \dots \} $.

{\bf Proof.} First we consider the case $q > 2$. In what follows we use the relation
  \begin{equation} 
      \frac{1}{3} < z_0 = z_0(p,q) < 0.4,
           \label{5.2}
  \end{equation}
  for all $p  > 0$ and  $q > 2$. Indeed, it follows from   
  relations (\ref{A.13a}), (\ref{A.14}) and (\ref{A.15}) given at Appendix that 
 \begin{equation} 
       \frac{1}{3} < z_0 = z_0(p,q) < z_{*}(q) \le z_{*}(3) = \frac{3 - \sqrt{5}}{2} \approx 0,382,
            \label{5.2a}
  \end{equation}
for all $p  > 0$ and  $q \ge 3$. Thus, relation (\ref{5.2}) is correct.

In what follows we use the following splitting
\begin{eqnarray} 
&&y = y_1 y_2 y_3,  \label{5.3} \\
 &y_1 =  \left[\frac{1+ p}{1+ p(1-z_0^q)}\right]^{2/q}, \quad   
  &y_2 =   (1 - z_0)^{2}, \quad     y_3 = \sqrt{\mathcal{B}(z_0)}, \label{5.4}
\end{eqnarray}
where $\mathcal{B}(z) = \frac{3}{2} q - \frac{2 (q - 2 ) z}{(1-z)^2}$.

For $y_1$ we obtain from (\ref{5.2})
\begin{equation} 
       y_1 = y_1(p,q) = \left[\frac{1}{1 - \frac{p}{p+1} z_0^q}\right]^{2/q} < 
         \left[\frac{1}{1 -  z_0^q}\right]^{2/q} < \left[\frac{1}{1 -  (0.4)^q}\right]^{2/q},
           \label{5.5}
  \end{equation}
for all $p > 0$ and $q > 2$.
Now, we use the following fact about the function
\begin{equation} 
      \tilde{f}(q) =  \left[\frac{1}{1 -  u^q}\right]^{2/q},
           \label{5.6}
  \end{equation}
where $0 < u < 1$ and $q  > 0$. Namely, the function $\tilde{f}(q)$ 
is monotonicall decreasing in $(0, + \infty)$. This follows 
from the relation 
\begin{equation} 
    \frac{d \tilde{f}(q)}{dq} = \tilde{f}(q) \frac{2}{q^2 (1-x)} 
    \left[ (1- x)\ln(1-x) + x \ln x  \right] < 0,
           \label{5.7}
  \end{equation}
where $x = u^q$ and $0 < x < 1$. This fact imlies for $u = 0.4$
the following bound
\begin{equation} 
       y_1 = y_1(p,q) < \left[\frac{1}{1 -  (0.4)^q}\right]^{2/q} 
       \le \left[\frac{1}{1 -  (0.4)^3}\right]^{2/3} \approx 1.04507975.
           \label{5.8}
  \end{equation}
for all $p > 0$ and  $q  \ge 3$. Hence, we get 
\begin{equation} 
       y_1  = y_1(p,q) <  1.0451,     \label{5.9}
  \end{equation}
for all $p > 0$ and  $q > 2$.

For $y_2$ we obtain from  (\ref{5.2})
\begin{equation} 
       y_2  = y_2(p,q) = (1 - z_0)^{2} <  \frac{4}{9},     \label{5.10}
  \end{equation}
for all $p > 0$,  $q > 2$.

The last bound 
\begin{equation} 
       y_3  = y_3(p,q)   =  \sqrt{\mathcal{B}(z_0)}
        < \sqrt{\mathcal{B}(1/3)} = \sqrt{3},     \label{5.11}
  \end{equation}
is also valid  for all $p > 0$ and  $q > 2$.
It follows from monotonical decreasing of the function $\mathcal{B}(z)$
in $(0,1)$ and $1/3 < z_0 < z_{*} < z_1$.  Here $\mathcal{B}(z) > 0$ for 
$z \in (0,z_1)$ and $z_{*} = z_{*}(q)$, $z_1 = z_1(q)$ are defined in Appendix.  

Plugging the bounds (\ref{5.9}), (\ref{5.10}), (\ref{5.11}) into (\ref{5.3})
we find 
\begin{equation} 
       y  = y(p,q) <  1.0451 \times (4/9) \times \sqrt{3} \approx 0.804518,     \label{5.12}
  \end{equation}
and hence
\begin{equation} 
       y  = y(p,q) <  0.80452 < 1,     \label{5.13}
  \end{equation}
for all $p > 0$,  $q > 2$.

\begin{figure}[h!]
	\begin{center}
		\includegraphics[width=0.66\linewidth]{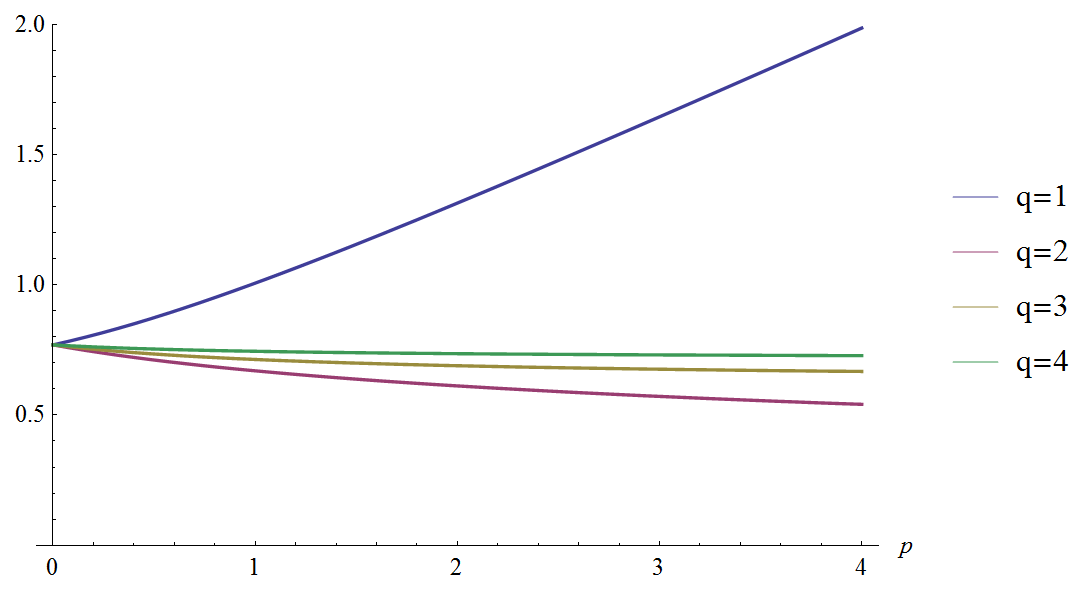}
		\caption{The graphical representation of the function  $y(p,q)$ for  $q = 1,2,3,4$}
		\label{yplot}
	\end{center}
 \end{figure}

This result can be illustrated by a numerical plot of the function $y(p,q)$ for a particular set of values of $q$, depicted on Fig.~\ref{yplot}. For $q=2$ the validity of Proposition 2 was verified numerically.

We note that recently, some examples of the violation of the Hod conjecture have been discussed for certain black hole solutions in supergravity and other theories  \cite{41}.

{\bf Remark}. Let us comment also on the case $q = 1$ which gives us the 
Reissner-Nordstr\"om  metric. It may be readily verified that in this case the inequality (\ref{5.1})
is not satisfied for all values of $p$: it is valid only for $0 < p < p_{cr}$, where $ p_{cr} $
is some critical value  of parameter $p$ \cite{MBI}. As it was pointed out in \cite{MBI} 
the violation of the Hod inequality in the eikonal regime for certain $p$ (and $n=0$)
 does not close the possibility for the obeying this relation for exact values of QNM for certain  
 $l = 0, 1, 2, \dots$ and all values of parameter $p$.

\section{Conclusion}

Here we have studied a non-extremal black hole solutions in a 4-dimensional gravitational model with anisotropic fluid proposed in Ref.~\cite{DIM}. The equations of state for the fluid  (\ref{0.4}) contains a parameter $q$ which is natural number  $q =1,2,3 \dots$. 
We have outlined  the global structure  of solutions under consideration: 
for odd $q = 2k + 1$  the Carter-Penrose diagram is coinciding with that of  Reissner-Nordstr\"om  
metric (the case of time-like singularity hidden by two horizons) while for  even $q = 2k$ it is coinciding with that of  Schwarzschild metric (the case of space-like singularity hidden by one horizon). 
For $q = 1$ the metric of the  solution  \cite{DIM} coincides with the metric  of   the Reissner-Nordstr\"om solution while in the limit  $q = + \infty$, we get the metric of the Schwarzschild solution.  
We have also presented certain physical parameters corresponding to BH solutions: gravitational mass $M$, Hawking temperature, black hole area entropy. 

We  have examined the solutions  to  massless Klein-Fock-Gordon equation in the background of our static BH metric for given $q =1,2,3 \dots$.  . By using the  tortoise coordinate  we have reduced this  equation to  radial one governed by  certain effective potential. This potential contains the parameters of solution such as $P >0$, $\mu >0$, natural parameter $q$ and also $l$ which is the multipole quantum number, $ l = 0, 1,\dots $.  

Here we have studied  the eikonal part of the effective potential for large $l$ and have found  a master equation for the value $z_0 = 1 - 2\mu / r_0$,  where  $r_0$  is the value of the radial coordinate (radius) $r_0$  corresponding to the maximum of the eikonal part of the effective potential. By using the maximum value of (the eikonal part) of the effective potential $\mathcal{V}_0$ and $r_0$, we have calculated the cyclic frequencies of the QNMs  in the eikonal approximation up to solution of the master equation  in   $z_0$.    Since the master equation  is an algebraic equation of order $q+1$ in $z_0$ we were able to find analytical exact solutions for $q = 1,2,3$. For obtained values of  eikonal  QNMs we have      also  considered   special cases $q = 1,2,3$  and a limiting cases $q =+\infty$. For  $q = 1$ our (eikonal) relations are  compatible with the well-known result for Reissner-Nordstr\"om solution \cite{AndOn} 
(for $n=0$), while for $q =+\infty$ they in an agreement with the well-known result for  the Schwarzschild solution \cite{BlMash}.    

We have also tested the validity of the Hod conjecture for  our solutions by considering QNMs  (eikonal) frequences with the lowest value of the overtone number $n=0$. 
We have shown that  the Hod conjecture is valid  in the range of $q > 1$.  This assumption  is valid for these values of  $q > 1$ since it is supported by  examples of states with large enough values of the multipole number $l$.

We note, that the  results obtained  here for eikonal QMN modes of test massless (non-charged) scalar field are also valid for  some other test fields, e.g. for electromagnetic one. This may be considered (by product) in a separate publication. (The results of Refs. \cite{46,47} may be also used in future work.)

\renewcommand{\theequation}{\Alph{section}.\arabic{equation}}
\renewcommand{\thesection}{\Alph{section}}
\setcounter{section}{0}

\section{Appendix}

\renewcommand{\thesubsection}{\Alph{section}}

Here we prove the Proposition 1. Since the extremality condition (\ref{4.6V})
for the effective potential  $\mathcal{V}$ 
($r > 2 \mu$) is equivalent to the master equation (\ref{4.6m})
  ($z = 1- 2\mu/r$), and the second derivative $\mathcal{V}_0''$ at the 
  point of extremum is given by relation (\ref{4.V2}) with $\mathcal{V}_0 > 0$
  (see (\ref{4.9V})), the Proposition 1 is equivalent to the following
  Lemma.
  
{ \bf Lemma}. {\em For any $p > 0$ and $q \in {\bf N} = \{1,2, 3 \dots \}$,  the 
 master equation 
    \begin{equation} \label{A.1}
     p z^{q+ 1} - 3 p z^{q} + (1+ p)(3z - 1) = 0
     \end{equation}
has only one solution $z_0 = z_0(p,q)$, belonging to interval $(0,1)$. 
 This solution obeys  the inequality  
  \begin{equation} \label{A.2}
      \mathcal{B}(z_0) = \frac{3}{2} q - \frac{2 (q - 2 ) z_0}{(1-z_0)^2} > 0
   \end{equation}
   for all  $p > 0$ and $q \in {\bf N}$.}

\begin{figure}[!h]
	\begin{center}
		\includegraphics[width=0.8\linewidth]{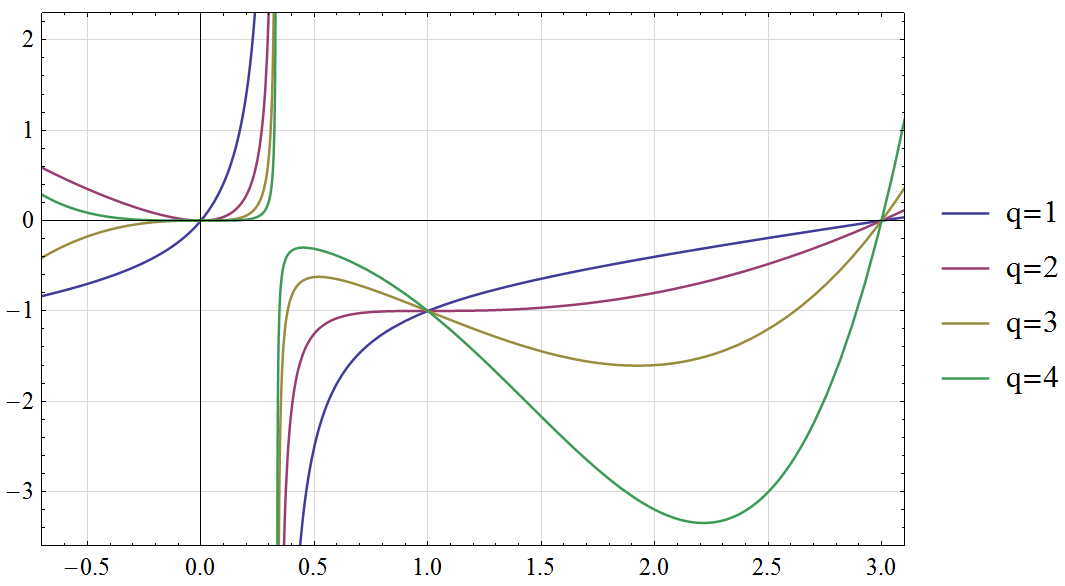}
		\caption{The graphical representation of the functions $F(z) =  
						F(z,q)$ for  $q = 1,2,3,4$.}
		\label{rfig:11}
	\end{center}
 \end{figure}

 {\bf Proof}.  Since $z = 1/3$ is not a solution
  to eq. (\ref{A.1}) we present the master equation in the following form
  \begin{equation} \label{A.3}
        F(z) = F(z,q) = z^q \frac{z-3}{3z-1} = - b = -1 - \frac{1}{p} < 0,
  \end{equation}
$p > 0$.  The functions $F(z) = F(z,q)$,  $q = 1,2,3,4$, are presented at Fig.~\ref{rfig:11}.
It follows from the definition (\ref{A.3}) that 
\begin{equation} \label{A.4}
        F(z,q) > 0,
  \end{equation}
for $z \in (0,1/3)$, $q \in {\bf N}$ and
\begin{equation} \label{A.5}
        \lim_{z \to 1/3 \pm 0} F(z,q) = \mp \infty,
  \end{equation}

\begin{equation} \label{A.6}
        \lim_{z \to 1 - 0} F(z,q) = -1,
  \end{equation}
for all $q \in {\bf N}$. Hence the seminterval $(0,1/3]$ should be excluded in our search
the solution to Eq. (\ref{A.3}). 

Let us analyze behavior of the function $F(z) = F(z,q)$ 
for $z \in (1/3,1)$ and fixed $q \in {\bf N} = \{1,2, 3 \dots \}$.   
The first derivative reads 
    \begin{equation} \label{A.7}
            \frac{dF(z)}{dz} = \frac{\partial F(z,q)}{\partial z} 
             = z^{q-1} \frac{[3q z^2 + (8 - 10q)z + 3q]}{(3z-1)^2}.
      \end{equation}
For $q = 1,2$, we have $\frac{dF(z)}{dz} > 0$ for $z \in (1/3,1)$ 
and hence the function $F(z)$ is monotonically increasing from $ - \infty$ to $-1$, when 
 $z \in (1/3,1)$. By applying the Intermediate Value Theorem 
 to our continuous monotonically increasing function $F(z) = F(z,q)$,
 $q=1,2$, we get that for any $p > 0$ there exist unique $z_0(p,q) \in (0,1)$,
 with $z_0(p,q) > 1/3$,  which obeys eq. (\ref{A.1}). 
 \footnotetext[1]{We remind that the Intermediate Value Theorem states that if $F$ is a continuous function defined on the interval $[a, b]$, then it takes on any given value between $F(a)$ and $F(b)$ at some point of this interval.}
 Inequality (\ref{A.2}) is obviously satisfied for $q=1,2$. 
 That means that the Lemma is valid for $q=1,2$.   

Now we consider the case $q > 2$. From (\ref{A.7}) 
we obtain that the there exists a unique point of extremum
of the function $F(z,q)$ in the interval  $(1/3,1)$   
 \begin{equation} \label{A.8}
    z_1 = z_1(q) = \frac{10q - 8 - \sqrt{(16q - 8)(4q -8)}}{6q},
  \end{equation}
$1/3 < z_1(q) <1$, which is the first root of the quadratic equation
$3q z^2 + (8 - 10q)z + 3q = 0$. The second root $z_2(q) = 1/z_1(q) \in (1,3)$
is irrelevant for our consideration.

The calculations give us: $z_1(3) = (11 - 2 \sqrt{10})/9 \approx 0,5195$, 
 $z_1(4) = (4 -  \sqrt{7})/3 \approx 0,4514$,  $z_1(5) = (7 -  2\sqrt{6})/5 \approx 0,4202$
 and  $F(z_1(3)) \approx -0,6227$, $F(z_1(4)) \approx -0,2987$, $F(z_1(5)) \approx -0,1297$. 
 We note that 
  \begin{equation} 
           \label{A.14a}
              z_{1}(q+1) < z_{1}(q),
        \end{equation}
 for all $q > 2$. This follows from monotonical decreasing of the function 
  $z_{1}(q)$ for $q > 2$, since  $z_1(q) = 1/z_2(q)$ and  
  \begin{equation} \label{A.14b}
      z_2(q) =   \frac{10 - 8/q +  \sqrt{(16 - 8/q)(4 -8/q)}}{6},
    \end{equation}
  is monotically increasing in $q$ for $q > 2$. 

It may be verified that
 \begin{equation} \label{A.9}
    z_1(q) \to \frac{1}{3}, \qquad F(z_1(q)) \to 0, 
  \end{equation} 
for $q \to + \infty$. Indeed, it follows from  (\ref{A.8}) 
that 
  \begin{equation} \label{A.10}
    z_1(q) =  \frac{1}{3} +  \frac{1}{3q} + O(q^{-2}), 
  \end{equation} 
  and
 \begin{equation} \label{A.11}
    F(z_1(q))  \sim  \frac{1}{3^q} \left(1+ \frac{1}{q} \right)^q \left(- \frac{8}{3} \right) q
     \sim  - \frac{8 e}{3^{q+1}} q \to 0
 \end{equation}
as $q \to + \infty$.

The function $F(z) = F(z,q)$ (for $q >2$) is monotonically increasing in the interval $(1/3,z_1)$, since
$\frac{dF(z)}{dz} > 0$ in this interval, see (\ref{A.7}), while it is  
monotonically decreasing in the interval $(z_1, 1)$ due to inequality 
$\frac{dF(z)}{dz} < 0$ which is valid there. Hence we get
       \begin{equation} \label{A.12}
          F(z_1(q),q) >  F(z,q) > F(1,q) = -1
       \end{equation}
 for all $z \in  (z_1, 1)$ and $q > 2$. This implies that the semi-interval $[z_1(q),1)$ 
 should be excluded in our search of solution to equation (\ref{A.3}) for a given $q > 2$. 
 Thus, we restrict our consideration to  $z \in (1/3, z_1(q))$.

 Let us define  $ z_{*}(q) \in (1/3, z_1(q)) $,  which obeys 
 the following equation 
           \begin{equation} \label{A.13}
                    F(z_{*}(q),q) = -1,
           \end{equation}
$q > 2$. By applying the Intermediate Value Theorem for a continuos 
monotonically increasing function  $F(z(q),q)$ defined on 
$(1/3, z_1(q))$ and using  (\ref{A.5}) and (\ref{A.12}) one
can readily prove that such point does exist and is unique  for any $q >2$. 

The calculations give us 
\begin{equation} 
  \label{A.13a}
  z_{*}(3) = \frac{3 - \sqrt{5}}{2} \approx 0,382, \quad
  z_{*}(4) \approx 0,346, \quad z_{*}(5) \approx 0,337.
\end{equation}

It may be proved that  
       \begin{equation} 
          \label{A.14}
             z_{*}(q+1) < z_{*}(q),
       \end{equation}
for any natural $q > 2$. Indeed, if we suppose that $z_{*}(q+1) \ge z_{*}(q)$ for some $q$
we get from monotonical increasing of the function $F(z,q+1)$ in $(1/3, z_1(q+1))$ 
and obvious inequality $F(z,q + 1) > F(z,q)$ for $z \in (1/3,1)$ that 
$$-1 = F(z_{*}(q+1),q +1 ) \ge  F(z_{*}(q),q + 1) > F(z_{*}(q),q) = -1 $$
and hence we come to a contradiction. Thus, the chain of inequalities  (\ref{A.14}) is
correct.  

Now we return to our original equation (\ref{A.3}). From  monotonical increasing of the function 
$F(z,q)$ in $(1/3, z_1(q))$ we get that $F(z) \ge F(z_{*}(q)) = -1 $ 
for $z \in [ z_{*}(q), z_1(q))$ and hence the semi-interval 
$[ z_{*}(q), z_1(q))$ should be excluded for our consideration of (\ref{A.3}).
By applying once more  the Intermediate Value Theorem for a continuos 
monotonically increasing function  $F(z(q),q)$ defined on 
$(1/3, z_{*}(q))$ and using  (\ref{A.5}) and (\ref{A.13}) we
can find that the point $z_0$ which  obeys the equation (\ref{A.3}) does exist, 
belongs to $(1/3, z_{*}(q))$ and is unique  for any $q > 2$ and  $p > 0$. 
We denote this point as $z_0 = z_0(p,q)$. 
Thus, we have 
\begin{equation} 
          \label{A.15}
       1/3 < z_0(p,q) < z_{*}(q) < z_{1}(q),
       \end{equation}
  for all $q > 2$ and  $p > 0$.                                         
  It follows from (\ref{A.9}) and  (\ref{A.15})
 \begin{equation} \label{A.16}
     z_0(p,q) \to \frac{1}{3},  
   \end{equation} 
 as $q \to + \infty$ uniformly in $p \in (0, + \infty)$.                                    

We note that one can present the solution as 
  \begin{equation} \label{A.17}
       z_0(p,q) = F_q^{-1} \left( -1 - \frac{1}{p} \right),  
     \end{equation} 
 where   $F_q^{-1}$ is the  function which is inverse to the function 
 $F_q : (1/3,z_{*}(q)) \longrightarrow (-\infty, -1)$, defined as
 $F_q (z) = F(z,q)$. The function $F_q^{-1}$ is a continuos  and 
 monotonically increasing one (due to a proper theorem on inverse function). 
 It may be readily verified that                  
 \begin{equation} \label{A.18a}
      \lim_{p \to + \infty} z_0(p,q) = z_{*}(q),  
 \end{equation}
and 
  \begin{equation} \label{A.18b}
        \lim_{p \to + 0} z_0(p,q) = 1/3.  
   \end{equation}

Thus, the first part of the Lemma is proved for all $q \in {\bf N}$. 
Now, we should prove the second part of the Lemma for $q >2$ 
(for $q=1,2$ it was checked above). Let us consider the function
  \begin{equation} \label{A.19}
      \mathcal{B}(z) = \frac{3}{2} q - \frac{2 (q - 2 ) z}{(1-z)^2}
   \end{equation}  
for $z \in (0,1)$ and $q = 3,4, \dots$. We get  
\begin{equation} \label{A.20}
      \mathcal{B}(z) = \frac{3q z^2 + (8 - 10q)z + 3q}{2(1-z)^2} =
      \frac{3q (z -z_1(q))(z -z_2(q))}{2(1-z)^2},
   \end{equation}  
where $z_1(q) < 1$ and $z_2(q) > 1$ are defined by relations (\ref{A.8}) 
and  (\ref{A.14b}), respectively. We find that 
$\mathcal{B}(z) >0 $ for all $z \in (0, z_1(q))$ and hence
for $z = z_0(p,q)$ with $q > 2$ and $p > 0$.  
We remind that  $1/3 < z_0(p,q) < z_{*}(q) < z_1(q)$ 
for all $q > 2$ and $p > 0$. Thus, the inequality (\ref{A.2})
is satisfied. The Lemma is proved.

%\begin{center}
 {\bf Acknowledgments}
%\end{center}

This paper has been supported by the RUDN University Strategic Academic Leadership Program (recipients: V.D.I. - mathematical model development and  S.V.B. - simulation model development).  
The reported study was partially funded by RFBR, 
project number 19-02-00346 (recipients  S.V.B. and V.D.I. - physical model development).

\end{document}